\documentclass[smallextended]{svjour3}       % onecolumn (second format)
\smartqed  % flush right qed marks, e.g. at end of proof
\usepackage{graphicx}
\usepackage{amssymb}
\begin{document}

\title{On the ontology of spacetime%\thanks{Grants or other notes
%about the article that should go on the front page should be
%placed here. General acknowledgments should be placed at the end of the article.}
}
\subtitle{Substantivalism, relationism, eternalism, and emergence}

\titlerunning{Ontology of spacetime}        % if too long for running head

\author{Gustavo E. Romero}

%\authorrunning{Short form of author list} % if too long for running head

\institute{Instituto Argentino de Radioastronom{\'{i}}a (IAR, CCT La Plata, CONICET) \at
              C.C. No. 5, 1894, Villa Elisa, Buenos Aires, Argentina. \\
              Tel.: +54-221-482-4903 ext. 115\\
              Fax: +54-221-425-4909 ext 117\\
              \email{romero@iar-conicet.gov.ar}
}

\date{Received: date / Accepted: date}
% The correct dates will be entered by the editor

\maketitle

\begin{abstract}

I present a discussion of some issues in the ontology of spacetime. After a characterisation of the controversies among relationists, substantivalists, eternalists, and presentists,  I offer  a new argument for rejecting presentism, the doctrine that only present objects exist. Then, I outline and defend a form of spacetime realism that I call event substantivalism. I propose an ontological theory for the emergence of spacetime from more basic entities (timeless and spaceless `events'). Finally, I argue that a relational theory of pre-geometric entities can give rise to substantival spacetime in such a way that relationism and substantivalism are not necessarily opposed positions, but rather complementary. In an appendix I give axiomatic formulations of my ontological views.

% \PACS{ 01.70.+w  \and 04.20.Gz }
% \subclass{MSC code1 \and MSC code2 \and more}
\end{abstract}
\vspace{0.3cm}

\begin{quotation}
\begin{flushright}
%$\varphi\'\upsilon\sigma\iota\zeta$ $\kappa\rho\'\upsilon\pi\tau\epsilon\sigma\theta\alpha\iota$ $\varphi\iota\lambda\epsilon\~\iota$.\\
For there neither is nor will be anything else besides what is, since Fate
has fettered it to be whole and changeless.\\[0.5cm]
{\sl Parmenides\footnote{Fragment 8. From the translation in G. S. Kirk, J. E. Raven, and M.
Schofield, The Presocratic Philosophers, 2nd ed. (Cambridge: Cambridge University Press,
1983), 252.}.}
\end{flushright}
\end{quotation}  

\section{Introduction}
\label{intro}

Discussions and controversies about the nature of space and time in Western thought can be traced to the early Pre-Socratic philosophers (Graham  2006, Jammer 2012,  Romero 2012). The position of Aristotle, who understood time as a measure of motion, and its contrast with the Platonic view, shaped the ontological controversy of the Hellenistic period, the Late Antiquity, and even the Middle Ages (see  Sorabji 1983). It was not, however, until the development of Newtonian physics and the Leibniz-Clarke correspondence (Leibniz and Clarke 2000), that what is now called relationism-substantivalism debate clearly emerged. Crudely, relationism is the metaphysical doctrine that space and time are not material entities existing independently of physical objects. Space and time emerge, according to this view, as a complex of relations among things and their changes. There are spatial and temporal relations among the constituents of the world, but not space and time in themselves. On the contrary, substantivalism is committed to the independent existence of space and time, which are considered as material substances or even as things or entities.  Substantivalism is ontological realism about space and time. The debate between both parties went on during more than 300 years (see the books by Jammer 2012 and Friedman 1983 for arguments supporting both positions). 

With the advent of the concept of spacetime (Minkowski 1908) and the General Theory of Relativity (Einstein 1915), the debate underwent such changes in the meaning of the original terms of both positions that some authors, as Rynasiewicz (1996), claimed that the whole issue was outmoded and ill-directed. Hoefer (1998) has argued, convincingly, that although some aspects of the classical debate might dissolve in the new context, the dispute is based on a genuine ontological problem and the debate goes on. I agree. I maintain, however, that the current ontological discussion cannot ignore the related issue of the eternalism-presentism-growing block universe. In this paper I shall offer a view of the topic in which a kind of substantivalism, relationism, and eternalism can coexist on the basis of emergentism, the doctrine that qualitative systemic properties arise from more basic ontological levels devoid of such properties. The mechanisms that enforce emergence are composition and interaction. I hold that there is a level for each of the three ontological positions to be considered as a good option for a description of the way the world is. 

In what follows, I first give a characterisation of the main concepts I am going to discuss and then I place the debate in the context of General Relativity and spacetime ontology. Next, I present a new argument for rejecting presentism, the doctrine that only the present time exists. In my opinion, this is the only of the four ontological views that is completely inconsistent with modern science. The remaining of the paper is devoted to outline my ontological position about the existence of spacetime. Technical details go to the Appendix, so the bulk of the discussion is apt for a broad readership.        

\section{The controversy}
\label{sec:1}

The traditional substantivalism--relationism debate was reshaped by the introduction of General Relativity in 1915. The changing views of Einstein himself on ontological matters helped to generate much confusion in the early interpretations of the theory. Einstein was originally  motivated in part by a Mach-inspired relationism (see his debate with de Sitter about the impossibility of empty spacetime, Smeenk 2014). Then, he shifted to a kind of ``ether substantivalism'' after 1918 (Einstein 1920, Hoefer 1998) to end espousing a block universe {\it \'a la} Weyl, after the early 1920s. He remained a spacetime realist and hardcore eternalist until the end of his life. He wrote to Vero and Bice Besso, referring to the death of his lifelong friend Michele Besso, just three weeks before his own death (F\"olsing 1998):\\

\begin{quotation}
Now he has preceded me a little by parting from this strange world. This means nothing. To us, believing physicists, the distinction between past, present, and future has only the significance of a stubborn illusion. \\
\end{quotation}

After a meeting with Einstein in 1950, Karl Popper wrote (Popper 2005, p.148):\\

\begin{quotation}
 I had met Einstein before my talk, first through Paul
Oppenheim, in whose house we were staying. And although I
was most reluctant to take up Einstein's time, he made me come
again. Altogether I met him three times. The main topic of our
conversation was indeterminism. I tried to persuade him to give
up his determinism, which amounted to the view that the world
was a four-dimensional Parmenidean block universe in which
change was a human illusion, or very nearly so. (He agreed that
this had been his view, and while discussing it I called him
“Parmenides”.) \\

\end{quotation}

But at the time of his debate with de Sitter (1917), Einstein wrote (Einstein 1918a, see also Smeenk 2014):\\

\begin{quotation}
 It would be unsatisfactory, in my opinion, if a world without
matter were possible. Rather, the $g^{\mu\nu}$-field should be fully determined by matter and not be able
to exist without the latter.\\
\end{quotation}

All these shifts  of ontological views by the founder of the theory contributed to create some confusion on the metaphysical trenches.  

The development, in the early 1920s, of dynamicist philosophical views of time by Bergson, Whitehead, and other non-scientific philosophers helped to resurrect presentism, the Augustinian view that only the present time exists and there is no future or past. Such a doctrine has a profound impact on theological issues and has been defended by Christian apologists (see, e.g. Craig 2008) but also by  scientific-oriented thinkers in later years (see, e.g. Bunge 1977). Substantivalism, relationism, eternalism, and presentism are all different ontological stances, although some of them are closely related. I offer some provisional definitions in order to make some semantical clarifications of importance for the subsequent discussions.

\begin{itemize}
\item {\em Spacetime substantivalism}\footnote{I follow the modern jargon and adopt the expression ``susbstantivalism'' instead of the more traditional (and less awkward) ``substantialism''. Unfortunately, philosophy and elegence of style not always go together. }: Spacetime is an entity endowed with physical properties. This position is clearly expressed by Einstein (1920). The exact nature of this entity is open to discussion. I shall defend an event substantivalism.\\

\item {\em Spacetime relationism}: Spacetime is not an entity that can exist independently of physical objects. Spacetime, instead, is a system of relations among different ontological items. The nature of these items is also open to discussion. I shall propose that there is a level where a form a relationism provides an adequate framework for current physics and that this is not in contraction with event substantivalism when the latter is applied to a different ontological level.  \\

\item {\em Eternalism (also known as Block Universe -- BU --)}: Present, past, and future moments (and hence events) exist. They form a 4-dimensional `block' of spacetime. Events are ordered by relations of earlier than, later than, or simultaneous with, one another. The relations among events are unchanging. Actually, they cannot change since time is one of the dimensions of the block. I have defended this position in Romero (2012 and 2013a). The reader is referred to these papers as well as to Peterson and Silberstein (2011) and references therein for further arguments.\\

\item {\em Presentism}: Only those events that take place in the present are real. This definition requires explanations of the terms `present' and `real'. Crisp (2003, 2007) offers elucidations. See also the mentioned paper by Craig (2008), and Mozersky (2011). Presentism has been subject to devastating criticisms since the early attacks by Smart (1964), Putnam (1967), and Stein (1968). See Saunders (2002), Petkov (2006), W\"uthrich (2010), Peterson and Silberstein (2011),  Romero (2012, 2015) for up-dated objections.

\end{itemize}   

A position intermediate between eternalism and presentism is the growing block universe proposal, strongly advocated in recent years by cosmologist G.R.S. Ellis. This position holds that past and present events exist, but future moments and events are not real. Reality would be a kind of growing 4-dimensional block, to which events are been added and go from non-existence to present and then to the past. The ultimate motivation for this proposal seems to be in some interpretation of quantum mechanics and a commitment with indeterminisim with respect to the future (e.g. Ellis and Rotham 2010, see also Broad 1923).  Several of the objections raised against presentism apply to the growing block universe, but I shall not make the case against it here. I leave the topic to a further communication.

\section{Some further objections against presentism}
\label{sec:2}

Most of the arguments against presentism are based on the Special Theory of Relativity; see the references cited in the previous section and the discussions in Craig and Smith (2008). Metaphysical arguments can be found, for instance, in Oaklander (2004) and Mellor (1998). Recently, several arguments based on General Relativity have been displayed against presentism. Romero and P\'erez (2014) have shown that the standard version of this doctrine is incompatible with the existence of black holes. In Romero (2015) I enumerate a number of additional objections based on General Relativity and modern cosmology. W\"uthrich (2010) discusses the problems and inconsistence of presentism when faced with Quantum Gravity. Here, I offer a new argument based on the existence of gravitational waves.  

The argument goes like this:\\

P$_1$. There are gravitational waves. \\

P$_2$. Gravitational waves have non-zero Weyl curvature.\\

P$_3$. Non-zero Weyl curvature is only possible in 4 or more dimensions.\\

P$_4$. Presentism is incompatible with a 4 dimensional world.\\

Then, presentism is false. \\  

The logic is sound, so let us review the premises of the argument to see whether there is some escape route for the presentist. The truth of P$_1$ is accepted by the vast majority of scientists working on gravitation. Gravitational waves are a basic prediction of General Relativity (Einstein 1916, 1918b).  Large gravitational wave detectors such as LIGO -- the Laser Interferometer Gravitational Wave Observatory -- have been constructed and are now under a process of  upgrading to meet the required sensitivity for wave detection. A space-based observatory, the Laser Interferometer Space Antenna or LISA, is currently under development by the European Space Agency (ESA). All this activity shows the confidence of the scientific community in the existence of gravitational waves. Indirect evidence for such existence is found from the  orbital decay of the binary pulsar  PSR B1913+16, discovered by Hulse and Taylor in 1974. The decay of the orbital period is in such accord with the predictions of General Relativity that both scientists were awarded the Nobel Prize in Physics 1993 (see, for instance Taylor and Weisberg 1982). So,  P$_1$ can be considered true within the context of our present knowledge of the universe. 

Premises P$_2$ and P$_3$ are necessarily true. Gravitational waves propagate in empty space, where the Einstein's field equations are reduced to:

$$ R_{ab}=0.$$

This expression means that the 10 coefficients of the Ricci tensor are identically null. But the full Riemann tensor\footnote{The Riemann tensor represents the curvature of spacetime. See Appendix A.1.} has 20 independent coefficients since is a rank 4 tensor. The remaining 10 components are expressed by the Weyl tensor. Then, since the gravitational waves are disturbances in the curvature, the Weyl tensor must be non-zero in their presence. If the dimensionality of the world were 3, as proposed by the presentists, the Riemann tensor would have only 6 independent components, and since in 3 dimensions the Einstein's equations in vacuum are reduced to 6, the Weyl tensor must vanish. Only in 4 or more dimensions gravity can propagate through empty spacetime (see Hobson et al. 2006, p.184, and Romero and Vila 2014, p. 19).     

Then, the presentist should either deny that presentism is incompatible with 4-dimensionalism or accept that presentism is false. But presentism is essentially the doctrine that things do not have temporal parts (Heller 1990). Any admission of temporal parts or time extension is tantamount to renounce to the basic claim of presentism: there are no future or past events. I conclude that presentism is utterly false.  I shall ignore this position in what remains of this article.  

\section{Event substantivalism and the emergence of things}
\label{sec:3}

In General Relativity, a specific model representing a sate of affairs is given by a triplet $\langle E,\; {\bf g},\; {\bf T} \rangle$, where $E$ is a 4-dimensional, real, differentiable pseudo-Riemannnian manifold, ${\bf g}$ is a (pesudo)metric tensor field of rank 2 defined on $E$, and ${\bf T}$ is another rank 2 tensor field representing the energy-momentum of the material entities accepted by the theory. Both tensor fields are related by the Einstein's field equations: $G_{ab}(g_{ab})=\kappa T_{ab}$, where $G_{ab}=R_{ab}-1/2\, g_{ab} R$ is the so-called Einstein's tensor, a function of the metric field and its second order derivatives.  Substantivalism is usually presented within the context of General Relativity in one of two types: manifold substantivalism and metric substantivalism (Hoefer 1996). The former is characterised as the view that the bare manifold represents spacetime (Earman and Norton 1987).   The latter, as the view that the metric field ${\bf g}$ represents substantival spacetime (Hoefer 1996). 

Two lines of attack on manifold substantivalism have been adopted by philosophers of spacetime and advocates of relationism: the hole argument and the `absence of structure' argument. The first one was originally conceived by Einstein, and resurrected by Earman and Norton (1987). The second, was presented by Mauldin (1988) and elaborated by Hoefer (1996). Let us briefly review them.    

Imagine a situation where the matter distribution is known everywhere outside some closed region of spacetime devoid of matter, the so-called hole. Then, the field equations along with the boundary conditions supposedly enable the metric field to be determined inside the hole. General covariance states that the laws of physics should take the same mathematical form in all reference frames. In two different frames, there are two solutions that have the same functional form and impose different spacetime geometries. If the coordinate systems in these frames\footnote{Notice that frames, contray to coordinate systems, are physical objects.} differ only after some time $t=0$, there are then two solutions; they have the same initial conditions but they impose different geometries after  $t=0$. This seems to imply a breakdown of determinism. Then, the manifold substantivalist should abandon determinism if she wants to remain a realist about spacetime points represented by the bare manifold (Norton 2014). Nothing observable, however, is made indeterminate by the hole argument, and hence the relationist escapes unscathed.

As noted by Hoefer (1996), the argument outlined above is not conclusive: without the premise that determinism is actually true, the argument has no force beyond the psychological conviction that determinism deserves a fighting chance. I see an additional problem: the substantivalist can claim that there are two types of determinism, namely, ontological and epistemological. The hole argument affects only the second type, since it concerns the predictions of the theory, not its ontological assumptions (i.e. that the points of the manifold represent events). But the existence of Cauchy horizons in many solutions of General Relativity is well established, so the hole argument adds essentially nothing to the epistemic problems of the theory. In any case, the hole argument prevents the univocal identification of bare points of the manifold with spacetime, not spacetime substantivalism. 

The second criticism of manifold substantivalism is based on the observation that the manifold, being just a topological structure, has not geometrical properties that are essential to any concept of spacetime (Mauldin 1988). In particular, without the metric field is not possible to distinguish spatial from temporal directions or to establish relations of `earlier than' and `simultaneous with'.  I agree. The manifold by itself has not structure enough as to provide a suitable representation of spacetime.  Hoefer (1996) concludes that the metric field ${\bf g}$ is a much better candidate to represent spacetime than the manifold. He observes that the metric field is clearly defined, and distinguishable from the matter field ${\bf T}$, which represents the contents of spacetime. The metric field cannot be null over finite regions of the manifold, contrary to other fields. If the metric field were just a physical field defined over spacetime, the geodetic motion would not be related with spacetime, but only with this field.  Hoefer also remarks that Einstein was of the opinion of that if the metric coefficients are removed, no spacetime survives the operation, since nothing is left, not even Minkoswki spacetime.  All spatio-temporal properties disappear with the metric. Based on his rejection of primitive identity for the points of the manifold, Hoefer proceeds to identify substantivalism with the claim that the metric represents spacetime and the manifold is a dispensable metaphysical burden.  

I concur with the opinion that the metric is indispensable for a representation of spacetime.  The metric provides all properties associated with spacetime. The manifold, however, do not seem dispensable to me. The whole spacetime is represented by the ordered pair $\langle E,\; {\bf g} \rangle$.  The elements of the pair represent different aspects of spacetime. The points of the manifold represent the existing events that form the world, and the metric represent their relational and structural properties. The identification of spacetime with a single element of the pair leads to problems. Instead, the representation of spacetime with  $\langle E,\; {\bf g} \rangle$ is in accord with the usual practice in science of representing entities with sets and properties with functions (Bunge 1974a, b). It might be argued, as Hoefer (1996) does, that spacetime points have no duration, and hence no trajectories in time, and they do not interact in any way with each other or with physical objects or fields, so it would be weird to assign them any kind of independent existence. My answer to this complaint is that of course points do not interact: they are the elements of the manifold that represent events. Events form the ontological substratum, and they do not move nor interact: change and interaction emerge from their ordering. At the level of analysis of General Relativity, events do not need to satisfy primitive identity neither. Only at a pre-geometric level events can be differentiated by a single property, their potential to generate further events (see Section \ref{sec:5} below). At the level at which General Relativity is valid, events do not need to be differentiated and it is this very fact that allows us to represent them by a manifold plus the metric. There is then no problem at all with embracing Leibniz Principle (i.e. diffeomorphic spacetime models represent the same physical situation). We can actually define a spacetime model as an equivalence class of ordered pairs  $\{\langle M,\; {\bf g} \rangle\}$ related by a diffeomorphism. In this class, the manifold provides the global topological properties and the continuum substratum for the definition of the metric structure. The representation of spacetime appears, therefore, as the large number limit of an ontology of basic timeless and spaceless events that can be identified only at a more basic ontological level. 

The ontological operation of composition `$\circ$' of events is a binary relation that goes from pairs of events to events. If $E$ is a set of events, and $e_i, \;i=1,...,n \in E$ represent individual events, then $\circ: E\times \;E \rightarrow E$ is  characterised by the following postulates:\\

\begin{itemize}
%\noindent
\item P$_1$. $(\forall e)_E\;(e \circ e = e)$.

\item P$_2$. $(\forall e_1)_{E}  (\forall e_2)_{E}\; (e_1 \circ e_2 \in E)$.

\item P$_3$. $(\forall e_1)_{E}  (\forall e_2)_{E}\; (e_1 \circ e_2\neq e_2\circ e_1)$.\\
      
\end{itemize}

We can introduce some definitions:\\

\begin{itemize}

\item D$_1$. An event ${e_{1}} \in E $ is composite $ \Leftrightarrow  \left(\exists {e_{2}}, {e_{3}}\right)_{E} \left( {e_{1}} ={e_{2}} \circ {e_{3}} \right) $.

\item D$_2$. An event ${e_{1}} \in E $ is basic $ \Leftrightarrow \; \neg \left(\exists {e_{2}}, {e_{3}}\right)_{E} \left({e_{1}} ={e_{2}} \circ {e_{3}} \right)$.

\item D$_3$.  $ {e_{1}}\subset {e_{2}}\ \Leftrightarrow {e_{1}} \circ {e_{2}} = {e_{2}}\ $ (${e_{1}}$ is part of ${e_{2}}\ \Leftrightarrow {e_{1}} \circ {e_{2}} = {e_{2}} $) .

\item D$_4$.  $ \textrm{Comp}({e}) \equiv\{{e_{i}}\in E \;|\; {e_{i}}\subset {e}\}$ is the composition of ${e}$.\\

\end{itemize}

Composition leads to a hierarchy of events, with basic events on the lower level and increasing complexity towards higher levels. Reality seems to be organised into levels, each one differentiated by qualitative, emerging properties. 
A level can be defined as a collection of events or things that share certain properties and are subject to some common laws that apply to all of them.
For example, all chemical processes share some properties and obey to chemical
laws, but do not have biological properties or are constrained by social
laws.

Higher levels have processes and things with some properties belonging to lower levels in
addition to specific ones. For instance,  I, a human being, have mass, experience chemical reactions,
and have biological functions. Conversely, an atom has not biological
properties. At some point of this hierarchy of events, things can be introduced as classes abstracted from large number of events (see Romero 2013a for formal definitions).  A thing-based ontology allows a simplification in the description of the higher levels of organisation of what is, essentially, an event ontology. 

The structure of the level system is given by (e.g. Bunge 2003b):
$$ {\cal L}=\langle L, \; <\rangle,   $$
where $L$ is a set of levels and $<$ is an ordering relation (precedence).
For any level $L_n$,  $L_n < L_{n+1}$ iff  $\forall (e)[e \in L_{n+1} \rightarrow$ Comp$(e ) \in L_n]$.

I differentiate at least 6 levels of organisation of reality. In order of increasing complexity, these are: ontological substratum $<$ physical $<$ chemical $<$ biological $<$ social $<$ technical. The first level is formed by basic events and precedes the emergence of physical things at the physical level.  Once events have multiplied and composed to a point where they can be represented with a continuum set, General Relativity can be formulated.  In the Appendix A.2, I present General Relativity as a physical theory that emerges from the basic ontological level. The first axiom, of ontological nature, postulates the existence of all events. Form the start, then, the theory can be labeled as `event substantivalism'. Spacetime is represented by the ordered pair $\langle E, \; {\bf g} \rangle$, not by the bare manifold $E$ or by the metric field ${\bf g}$. Spacetime is then an emerging thing from the collection of all events, that can be characterised as an individual endowed with properties (Romero 2012, 2013a).  

I close this section offering a brief new argument for spacetime substantivalism. It might be called a `thermodynamical' argument: 

\begin{itemize}
%\noindent
\item P$_1$. Only substantival existents can be heated. 

\item P$_2$. Spacetime can be heated.      
\end{itemize}

Then, spacetime has substantival existence. \\

The logic is clearly sound, so let us briefly discuss the premises. P$_1$ is a fundamental insight from physics. To heat something is to excite its internal degrees of freedom. It is impossible to to heat something that does not exist, because non-existents do not have internal microstructure. Regarding P$_2$, quantum field theory in curve spacetime clearly indicates that spacetime can be heated and the amount of radiation produced by it can be increased (for instance, by acceleration or gravitational collapse, e.g. Birrell and Davis 1982). I conclude that spacetime has substantival existence.

\section{Defending eternalism}
\label{sec:4}

The assumption that the collection of all events exists and is represented by a 4 dimensional differentiable real manifold, along with the metric structure of this manifold given by the field ${\bf g}$, leads to the doctrine we have define as `eternalism': past, present, and future events exists. In fact, the metric allows to define the separation of any two events, $ds^2(e_1,\;e_2)=g_{ab}dx^a dx^b$, with $d{\bf x}$ the differential 4-dimensional distance between $e_1$ and $e_2$. According to whether $ds^2=0$, $ds^2>0$, or $ds^2<0$, the events are considered `null', `time-like', or `space-like', respectively. In the first two cases the events might be (but not necessarily are) causally related and the temporal ordering cannot be reversed with a simple change of coordinates. In the case of space-tike events, on the contrary, there is no absolute temporal ordering, given the invariance of the theory with respect to the group of general coordinate transformations. Events that are future or past in some system, can be simultaneous in another. If someone claims that a couple of space-like events are present, she must accept that there are future and past events (since there will be always a coordinate transformation that render them future or past) or negate that existence is invariant under coordinate transformations. The latter seems to be an impossible step. The existence of future and past events, hence, is implied by substantivalism, i.e. any consistent substantivalist must be an eternalist. The converse is not true. 

The existence of space-like events cannot be denied by a presentist, since the existence of all events was assumed from the very beginning, when the existence of the referents of the the manifold $E$ was accepted in the formulation of General Relativity (Axiom {\bf P$_1$} in A.2).  The presentist can try to offer a suitable reformulation of General Relativity where all but present events are just convenient fictions, but it is difficult to see how this move will help her to escape from the argument from general covariance, since the `present' is defined as a moving hypersurface of space-like events. For the eternalist, instead, there is nothing dynamical associated with the `present': this is just a local relational property; every event is present to a person located at that moment and location. The same event is past or future to persons located in the future or the past of the event; there is no intrinsic `presentness' associated with individual events. All events exists on equal foot for the eternalist.  

The presentist can object that eternalism implies fatalism: the future is fixed and unchangeable. This objection seems to be the main motivation for the postulation of the growing block universe view. The presentist's universe, however, can be as fixed in regards to the future as the block universe of the eternalist. This is because the inevitability of an occurrence depends on the character of the physical laws. If the laws are deterministic, the future of the presentist is still nonexistent, but will exist in a determined way. So the argument can work only if the presentist can prove that ontological determinism is false. The usual move here is to turn to quantum mechanics. There is, however, no help to be found in quantum theory since it does not imply the fall of ontological determinism. Two quantum events can be related by some probability estimated from the deterministic evolution of dynamical objects of the theory (either operators or wave functions, depending on the formulation). Such a relation, from the point of view of the spacetime, is as fixed as any other relation between the events. There is no sudden change of probabilities: the probabilities are just a mathematical measure of the propensity of the some events to be related. Besides mathematical objects like probabilities do not change. In this sense, quantum probabilities are no special: the probability of a dice roll to yield a 3 is 1/6, both before and after the rolling (see Romero 2015, appendix). This does not make less ontologically determined the events of throwing the dice and getting the 3. There is no `collapse' of the wave function. Wave functions, mathematical objects in the Hilbert space, cannot `collapse' in any meaningful sense (Bunge 1967, 1973; Bergliaffa et al. 1993).  What can change is a quantum physical system, not the probability attributed to the event by quantum mechanics. The evolution of the system, when it interacts, is not unitarian and cannot be predicted by quantum mechanics. It must be studied by a quantum theory of measurements, where each case depends of the specific instrumental set up. 

I also want to emphasise that quantum mechanics is not a background independent theory: it is formulated on a previously assumed spacetime theory (Euclidean spacetime in the case of non-relativistic quantum mechanics, Minkowski and pseudo-Riemannian spacetimes in the cases of relativistic quantum mechanics and quantum field theory on curve space). Being a background dependent theory, quantum mechanics imports the ontological assumptions of its background (Rovelli 2004).  So, if we have assumed a substantivalist view of spacetime, eternalism, far from being ruled out by quantum mechanics, is assumed as well in accordance with the implication we saw above. 

The other standard argument against eternalism raised by presentists is that it cannot explain the human experience of time and passing. I have addressed this issue in Romero (2015) so I shall only mention here that modern neuroscience supports the idea that the ``passage of time'' is a construction resulting from the ordering of brain processes (P\"oppel 1988, Le Poidevin 2007, Eagleman 2009).

\section{Relationism before time}
\label{sec:5}

Event substantialism regarding spacetime does not preclude relationism at a more basic level. Relations among basic events, or `ontological atoms'\footnote{These basic events can be thought as some suitable re-interpretation of Leibniz monads (Leibniz 2005).}, can be the basis from which substantival spacetime emerges, in a similar way to how things emerge from spacetime events.  

The manifolds adopted in General Relativity to represent spacetime have a pseudo-Riemannian metric and are compact. A very important property of such manifolds is that they are compact if and only if every subset has at least one accumulation point. These points are defined as:\\

{\bf Definition.} Let $E$ be a topological space and $A$ a subset of $E$. A point $a\in A$ is called an {\em accumulation point} of $A$ if each neighbourhood of $a$ contains infinitely many points of $A$.\\

For compact Hausdorff spaces\footnote{A manifold $E$ is said to be Hausdorff if for any two distinct elements $x\in E$ and $y\in E$, there exist $O_{x}\subset E$ and $O_{y}\subset E$ such that $O_{x} \cap O_{y}=\emptyset$.}, every infinite subset $A$ of $E$ has at least one accumulation point in $E$. 
 
If we want to represent events at very small scale, the assumption of compactness must be abandoned. The reason is that any accumulation point implies an infinite energy density, since events have finite (but not arbitrarily small) energy, and energy is an additive property. In other words, spacetime must be discrete at the smallest scale\footnote{Arguments for discrete spacetime coming from physical considerations can be found, for instance, in Oriti (2014) and Dowker (2006). Also, notice that the thermodynamical argument for the existence of spacetime presented in Sect. \ref{sec:3} implies that there exists a microstructure of spacetime, namely:

\begin{itemize}
%\noindent
\item P$_1$. Spacetime has entropy. 

\item P$_2$. Only what has a microstructure has entropy.      
\end{itemize}

Then, spacetime has a microstructure. }.   

As far as we can decompose a given event $e\in E$ into more basic events, in such a way that $E$ can be approximated by a compact uncountable (non-denumerable) metric space, the continuum representation for the totality of events will work. But if there are atomic events, there will be a sub-space of $E$ that is countable (or denumerable if it is infinite) and ontologically basic. There is, in such a case, a discrete substratum underlying the continuum manifold. Since the quantum of action is given by the Planck constant, it is a reasonable hypothesis to assume that the atomic events occur at the Planck scale, $l_{\rm P}=\sqrt{\hbar G/c^{3}}$. If there are atomic events, their association would give rise to composed events (i.e. processes), and then to the continuum spacetime, which would be a large-scale emergent entity, absent at the more basic ontological level. This is similar to, for instance, considering the mind as a collection of complex processes of the brain, emerging from arrays of `mindless' neurons.

If this view is correct, then quantum gravity is a theory about relations among basic events and the ontological emergence of spacetime and gravity. Quantum gravity would be a theory so basic that it might well be considered as ontological rather than physical.   

The discrete spacetime ontological substratum can be formed by atomic timeless and spaceless events. These events have only one intrinsic property: energy, i.e. the capability to generate more events. The relational properties of basic events result in spatio-temporal properties of the collection of composed events.  It has been suggested by Bombelli et al. (1987) that basic events and their relations should be represented by a partially ordered set, also called a {\em poset}. It can be proved that the dimension, topology, differential structure, and metric of the manifold where a poset is embedded is determined by the poset structure (Malament 1977). If the order relation is interpreted as a causal relation, the posets are called {\em causal sets} (or {\em causets}). We do not need to make this distinction here.

A poset is a set $P$ with a partial order binary relation $\preceq$ that is reflexive, antisymmetric, transitive, and locally finite (in the sense that the cardinality of the poset is not infinite, and hence there are no accumulation points). It is the local finiteness condition that introduces spacetime discreteness. 

A given poset can be embedded into a Lorentzian manifold taking elements of the poset into points in the manifold such that the order relation of the poset matches the ordering of the manifold. A further criterion is needed, however, before the embedding is suitable. If, on average, the number of  poset elements mapped into a region of the manifold is proportional to the volume of the region, the embedding is said to be faithful. The poset is then called {\em manifold-like}

A poset can be generated by {\em sprinkling} points (events) from a Lorentzian manifold. By sprinkling points in proportion to the volume of the spacetime regions and using the causal order relations in the manifold to induce order relations between the sprinkled points, a poset can be produced such that (by construction)  be faithfully embedded into the manifold. 
To maintain Lorentz invariance, this sprinkling of points must be done randomly using a Poisson process. 

A {\em link} in a poset is a pair of elements $x,\; y \in P$ such that $x \prec y$ but with no $z \in 
P$ such that $x \prec z \prec y$. In other words, $x$ and $y$ represent directly linked events. A {\em chain} is a sequence of elements $x_0,\; x_1,\ldots,x_n$ such that $x_i \prec x_{i+1}$ for $i=0,\ldots,n-1$. The length of a chain is $n$, the number of relations used. A chain represents a process.

A geodesic between two poset elements can then be introduced as follows: a geodesic between two elements $x,\; y \in P$ is a chain consisting only of links such that $x_0 = x$ and $x_n = y$. The length of the chain, $n$, is maximal over all chains from $x$ to $y$. In general there will be more than one geodesic between two elements. The length of a geodesic should be directly proportional to the proper time along a time-like geodesic joining the two spacetime points if the embedding is faithful.

A major challenge is to recover a realistic spacetime structure starting from a numerable poset. This problem is sometimes called ``dynamics of causets''. A step in the direction of solving the problem  is a classical model in which elements are added according to probabilities. This model is known as classical sequential growth (CSG) dynamics (Rideout and Sorkin 2000). The classical sequential growth model is a way to generate posets by adding new elements one after another. Rules for how new elements are added are specified and, depending on the parameters in the model, different posets result. The direction of growth gives rise to time, which does not exist at the fundamental poset event level. 

Another challenge is to account for the remaining referents of General Relativity, namely, gravitating objects. I have already suggested that physical objects can be understood as clusters of processes\footnote{Events are understood by some authors as changes in material objects (e.g. Bunge 1977). This definition is correct only above certain level of composition, at which basic events are irrelevant. There is not problem of circularity, then, with the views presented here. One can even reserve the name ``event'' for the changes in things, and adopt ``monads'' or some other fancy name for what I call here ``basic events''.}, and hence they might emerge as inhomogeneities in the growing pattern of events (Romero 2013a). This conjecture is supported by the observation that whatever exists seems to have energy, and energy is just the capability to change (Bunge 2003a). The more numerous the bundle of events is, the larger the associated energy results. Physical things, objects endowed of energy, would be systems formed by clusters of events. In Appendix A.3 I present an outline of an axiomatics of this pre-geometric ontological theory.

\section{An ontology cozy for science}
\label{sec:6}

The current physical view of the world is a collection of quantum fields existing in spacetime. The interaction of these fields is local. The properties of spacetime are represented by what is usually interpreted as another physical field, the Lorentzian metric field defined on the continuum 4-dimensional manifold. This field represents both the geometrical properties of spacetime and the potential of gravity. This dual character makes it unique among all physical fields. The metric tensor field, contrarily to the others, is a classical field with infinite degrees of freedom and background independence. Background-independence is the property that the metric of spacetime is the solution of the dynamical equations of the theory. 

When standard quantisation techniques are applied to gravity, there appear infinitely many independent parameters needed to correctly define the theory. For a given choice of those parameters, one could make sense of the theory; but since it is not possible to carry out infinitely many experiments to fix the values of every parameter, a meaningful physical theory cannot be determined: gravity is perturbatively nonrenormalizable. The appearance of singularities in General Relativity, however, indicates that the theory is incomplete (e.g. Romero 2013b). Another hint that a quantum theory of gravity should emerge from a discretisation of spacetime itself comes from black holes. Quantum field theory in curved spacetime shows that the horizon of a black hole has entropy. But the horizon is just a region of spacetime. Spacetime, hence, has an associated entropy. A merely continuum spacetime, with its infinite number of degrees of freedom would have an infinitely large entropy. The finiteness of the black hole entropy, then, points to the existence of a discrete substratum for spacetime. 

There is another very important difference between the metric field ${\bf g}$ and the fields of the Standard Model of particle physics. The ten coefficients of metric do not represent a physical field, but a class of properties of a substantival entity: spacetime. It is then incorrect to attribute energy to  ${\bf g}$. Properties do not have energy, only substantival entities have (Bunge 1977). Attempts to construct a well-defined and conserved energy for the metric field fail, and only a (non-unique) pseudo-tensor can be constructed. The reason is that the geometrical properties of spacetime are always locally reduced to those of a flat Minkowskian manifold. Physically, we call this condition `the Equivalence Principle' (Einstein 1907). Energy should be attributed not to the metric, but to substantival spacetime itself. The energy content of spacetime is related to the number of basic events per unit of volume. This number is minimum for nearly flat spacetime, or when the volume is very small ($\sim l_{\rm P}^3$), but it is never zero. It is not possible to eliminate the energy of spacetime through a transformation of coordinates, in the way the metric field can be made locally Minkowskian; existence cannot be suppressed by a mere coordinate change. I suggest that the average minimum energy of spacetime is measured by the cosmological constant. If there is only one basic event in a Planck cubic volume, the energy of such event would be amazingly tiny: $\sim 10^{-91}$ eV.      

The ontological views I advocate in this paper are in good agreement with these physical considerations. First, spacetime has substantival existence. It can be formally represented by a continuum manifold equipped with a metric tensor field: $ {\rm ST}\hat{=}\left\langle E, {\bf g}\right\rangle$.  Second, the existence of spacetime implies the existence of events that are past, present, and future. Third, the metric field is not akin other physical fields; it represents the geometrical properties of spacetime and does not have independent existence. And forth, as all large scale entities, spacetime emerges from the composition of more basic existents, that I have called `basic events'. I suggest that these ontological views can provide an adequate philosophical background for physical research of gravity and cosmology, both classical and quantum.      

\section{Closing remarks}
\label{sec:7}

Undoubtedly, ontology by itself cannot offer a solution to the problems of quantum gravity. But this is not the task of ontology. What should be expected from ontological theories is a framework suitable for the development of scientific research, with no hidden assumptions or confusing terms; a clarification of the basic concepts of our most general theories about the world and its emergence. It is in this sense that I think that a scientifically informed ontology can pave the way for research through the elucidation of our ideas of space, time, and spacetime. The considerations presented in this article were aimed in such direction.

\section*{Appendix: Axiomatics}
\label{Appendix}

\subsection*{A.1 Basic definitions}
In this appendix I give some basic definitions used in the two axiomatisations that follows. 

The Einstein tensor is:
\begin{equation}
	G_{ab}\equiv R_{ab}-\frac{1}{2}R g_{ab},
\end{equation}
where $R_{ab}$ is the Ricci tensor formed from second derivatives of the metric and $R\equiv g^{ab}R_{ab}$ is the Ricci scalar. The geodetic equations for a test particle free in the gravitational field are:
\begin{equation}
	\frac{d^{2}x^{a}}{d\lambda^{2}}+ \Gamma^{a}_{bc}\frac{dx^{b}}{d\lambda}\frac{dx^{c}}{d\lambda}=0,
\end{equation}
with $\lambda$ an affine parameter and $\Gamma^{a}_{bc}$ the affine connection, given by:
\begin{equation}
\Gamma^{a}_{bc}=\frac{1}{2}g^{ad}(\partial_{b}g_{cd}+\partial_{c}g_{bd}-\partial_{d}g_{bc}).	
\end{equation}
  
The affine connection is not a tensor, but can be used to build a tensor that is directly associated with the curvature of spacetime: the Riemann tensor. The form of the Riemann tensor for an affine-connected manifold can be obtained through a coordinate transformation 
${x^{a}\rightarrow {\bar{x}^{a}}}$ that makes the affine connection  vanish everywhere, i.e.

\begin{equation}
	\bar{\Gamma}^{a}_{bc}(\bar{x})=0, \;\;\; \forall\; \bar{x},\; a,\;b,\; c.
\end{equation}

\vspace{0.2cm}

\noindent The coordinate system ${\bar{x}^{a}}$ exists if

\begin{equation}
\Gamma^{a}_{bd, c}-\Gamma^{a}_{bc, d} + \Gamma^{a}_{ec}\,\Gamma^{e}_{bd} - \Gamma^{a}_{de}\,\Gamma^{e}_{bc}=0
\label{R}
\end{equation}

\noindent for the affine connection $\Gamma^{a}_{bc}({x})$. The left hand side of Eq. (\ref{R}) is the Riemann tensor:
\begin{equation}
	R^{a}_{bcd}=\Gamma^{a}_{bd, c}-\Gamma^{a}_{bc, d} + \Gamma^{a}_{ec}\,\Gamma^{e}_{bd} - \Gamma^{a}_{de}\,\Gamma^{e}_{bc}.
\end{equation}
 
When $R^{a}_{bcd}=0$ the metric is flat, since its derivatives are zero. If 
$$\mbox{$K=R^{a}_{bcd}R^{bcd}_{a}>0$}$$
 the metric has a positive curvature. Sometimes it is said, incorrectly, that the Riemann tensor represents the gravitational field, since it only vanishes in the absence of fields. On the contrary, the affine connection can be set locally to zero by a transformation of coordinates. This fact, however, only reflects the equivalence principle: the gravitational effects can be suppressed in any locally free falling system. In other words, the tangent space to the manifold that represents spacetime is always Minkowskian.

\subsection*{A.2 Axiomatic ontology of spacetime}

The basic assumption of the ontological theory of spacetime I propose is:\\

{\em Spacetime is the emergent system of the ontological composition of all events}. \\

Events can be considered as primitives. They are characterised by the axiomatic formulation of the theory.  Since composition is not a formal operation but an ontological one, spacetime is neither a concept nor an abstraction, but an emergent entity. What I present here is, then, a substantival\footnote{An entity $x$ has subtantival existence iff $x$ interacts with some $y$, such that $y\neq x$.} ontological theory of spacetime. As any entity, spacetime can be represented by a concept. The usual representation of spacetime is given by a 4-dimensional real manifold $E$ equipped with a metric field $g_{ab}$:

$$  
{\rm ST}\hat{=}\left\langle E, g_{ab}\right\rangle.
$$

I insist: spacetime {is not} a manifold (i.e. a mathematical construct) but the ``totality'' of all events. A specific model of spacetime requires the specification of the source of the metric field. This is done through another field, called the ``energy-momentum'' tensor field $T_{ab}$. Hence, a model of spacetime is:

$$  
M_{\rm ST}=\left\langle E, g_{ab}, T_{ab}\right\rangle.
$$

The relation between both tensor fields is given by the field equations. The metric field specifies the geometry of spacetime. The energy-momentum field represents the potential of change (i.e. event generation and density) in spacetime. 

We can summarise all this through the following axioms. The axioms are divided into syntactic, if they refer to purely formal relations, ontological, if they refer to ontic objects, and semantic, if they refer to the relations of formal concepts with ontological ones. There are no physical axioms at this level.\\

The basis of primitive symbols\footnote{A primitive symbol is a symbol not defined explicitely in terms of other symbols.} of the theory is:

$$B_{\rm Ont} =\left\langle {\cal E}, \; E, \; \left\{{\bf g}\right\}, \;\left\{ {\bf T} \right\},  \;\left\{ {\bf f} \right\}, \;\Lambda,\;\kappa \right\rangle. $$

\begin{itemize}

\item ${\bf P1 - Ontological/ Semantic}.$ $\cal{E}$ is the collection of all events. Every member $e$ of  $\cal{E}$ denotes an event. 

\item ${\bf P2 - Syntactic}.$ $E$ is a $C^{\infty}$ differentiable, 4-dimensional, real pseudo-Riemannian manifold.

\item ${\bf P3 - Syntactic}. $ The metric structure of $E$ is given by a tensor field of rank 2, $g_{ab}$, in such a way that the differential 4-dimensional distance $ds$ between two events is: $$ds^{2}=g_{ab} dx^{a} dx^{b}.$$

\item ${\bf P4 - Syntactic}.$ The tangent space of $E$ at any point is Minkowskian, i.e. its metric is given by a symmetric tensor $\eta_{ab}$ of rank 2 and trace $-2$,

$$	\eta_{ab} = \left(\matrix{1 &0&0&0\cr 0&$-1$&0&0\cr
  0&0&$-1$&0 \cr 0&0&0&$-1$\cr}\right)$$.

\item ${\bf P5 - Syntactic}.$ The symmetry group of $E$ is the set of all 4-dimensional transformations $\left\{ {\bf f} \right\}$ among tangent spaces. 

\item ${\bf P6 - Syntactic}.$ $E$ is also equipped with a set of second rank tensor fields $\left\{ {\bf T} \right\}$.

\item ${\bf P7 - Semantic}.$  The elements of  $\left\{ {\bf T} \right\}$ represent a measure of the clustering of events.

\item ${\bf P8 - Ontological - inner \; structure}.$ The metric of $E$ is determined by a rank 2 tensor field $T_{ab}$ through the following  equations:

\begin{equation}
{\bf G}-{\bf g} \Lambda=\kappa {\bf T}, \label{Eq-Einstein2} 
\end{equation}

or 

\begin{equation}
G_{ab}-g_{ab}\Lambda=\kappa T_{ab}, \label{Eq-Einstein1} \\
\end{equation}

\noindent where $G_{ab}$ is the Einstein tensor. Both $\Lambda$ and $\kappa$ are constants.

\item ${\bf P9 - Semantic}.$ The elements of $E$ represent physical events.

\item ${\bf P10 - Semantic}.$ Spacetime is represented by an ordered pair $\left\langle E, \; g_{ab}\right\rangle$: $${\rm ST}\hat{=}\left\langle E, g_{ab}\right\rangle.$$

\item ${\bf P11 - Semantic}.$ A specific model of spacetime is given by: $$M_{{\rm ST}}=\left\langle E, g_{ab}, T_{ab}\right\rangle.$$

\end{itemize}

This theory characterise an entity that emerges from the composition of basic, timeless and spaceless events (see below). On the basis of this theory we can formulate a physical theory about how this entity, spacetime, interacts with other systems and the corresponding dynamical laws. Such a theory is General Relativity.  The axioms we should add to obtain General Relativity form our ontological theory are:

\begin{itemize}

 \item ${\bf A.1 - Semantic}$ The tensor field ${\bf T}$ represents the energy, momentum, and stress of any physical field defined on $E$.

\item ${\bf A.2 - Physical}$ $\Lambda$ is a constant that represents the energy density of spacetime in the absence of non-gravitational fields. The constant $\kappa$ represents the coupling of the gravitational field with the non-gravitational systems.

\item ${\bf A.3 - Semantic }$ $k=-8\pi G c^{-4}$, with $G$ the gravitational constant and $c$ the speed of light in vacuum. 

\end{itemize}

From $\bigwedge_{i=1}^{11} {\bf P_i} \; \wedge \; \bigwedge_{i=1}^3 {\bf A_i} $, all standard theorems of General Relativity follow (see Bunge 1967, Covarrubias 1993).  

\subsection*{A.3 Towards an axiomatic pre-geometry of spacetime}

The ontological, substantival theory of spacetime outlined above characterise an entity, spacetime, that is formed by events. If events are the basic constituents of spacetime, a constructive theory of spacetime can be proposed. In such a theory, spacetime emerges from timeless and spaceless events whereas metric properties and the internal spacetime structure are the result of the transition to large numbers of events  that allows to adopt a continuum description. The development of a theory of this class is the major goal of several approaches to quantum gravity. In what follows, I outline a minimum axiomatic system that might be useful as a guiding framework for such an enterprise.  

The basis of primitive symbols of the theory is:\\

$$B_{\rm Pre-Geom} =\left\langle {\cal E}_{\rm B}, \; E_{\rm B}, \; P, \; \preceq, \; {\cal W}, \;  l_{\rm P}, \; \circ \right\rangle. $$\\
 
Tentative axiomatic basis:\\

\begin{itemize}

\item ${\bf O1 - Ontological/Semantic}.$ ${\cal E}_{\rm B}$ is the collection of basic events. Every $x$ in ${\cal E}_{\rm B}$  denotes an event. 

\item ${\bf O2 - Syntactic/Semantic}.$ There is a set ${E}_{\rm B}$ such that every  $e \in {E}_{\rm B}$  denotes a basic event of ${\cal E}_{\rm B}$. 

\item ${\bf O3 - Syntactic}.$ There is a binary operation $\circ$ from $E_{\rm B} \times E_{\rm B}$ into a set $E^*$ that composes basic events into complex events (Def. Complex events: {\em processes}). 

\item ${\bf O4 - Syntactic}. $ There exists a partially ordered set $P\subset {E}_{\rm B}$ (poset) endowed with the ordering relation $\preceq$. 

\item ${\bf O5 - Syntactic}.$ The partial order binary relation $\preceq$ is:
\begin{itemize} 
\item Reflexive: For all $x \in P$, $ x \preceq x$.
\item Antisymmetric: For all $x,\; y\; \in P$, $x \preceq y \preceq x$ implies $x = y$.
\item Transitive: For all $x,\; y,\; z \in P$, $x \preceq y \preceq z$ implies $x \preceq z$.
\item Locally finite: For all $x,\; z \in P$, {\bf card} $(\{y \in C | x \preceq y \preceq z\}) < \infty$.
\end{itemize}

Here {\bf card} $(A)$ denotes the {\em cardinality} of the set $A$. Notice that $x \prec y$  if $x \preceq y$ and $x \neq y$.

\item ${\bf O6 - Ontological}.$ The elements of  $\cal{E}_{\rm B}$ have an extensive property called {\em energy} ${\cal W}(x): {E}_{\rm B} \rightarrow  \Re$. The larger ${\cal W}(x)$, the more numerous are the events that can be linked to $x$ by $\preceq$ in $E^*$. 

\item ${\bf O7 - Ontological}.$ If  Comp$(e)=\{e_1, e_2, ..., e_n\}$ then ${\cal W}(e)={\cal W}(e_1)+{\cal W}(e_2)+...+{\cal W}(e_n)$, where all $e_i$ are basic events. 

\item ${\bf O8 - Ontological}.$ If $E' \subset E_{\rm B}$ has $n$ elements, then\\

$${\cal W}(E')=\Sigma_{i=1}^n {\cal W}(e_i),\;\;\; e_i \in E'$$\\

\item ${\bf O9- Syntactic}.$ $E_{\rm B}$ is embedded in $E^*$ in such a way that $E^*$ preserves the internal structure of $E_{\rm B}$  given by the relation of precedence. 
 
\item ${\bf O10- Syntactic}.$ The set $E^*$ has a (pseudo) metric structure.

\item ${\bf O11- Syntactic}.$ $E^*$ can be extended into a continuous, differentiable pseudo-Riemaniann 4-dimensional manifold $E$. 

\item ${\bf O12- Ontological}.$  The energy density is $\rho={\cal W}(E')/V$, where $V$ is the volume of a region $E'$ in $E$. This energy density forms a component of a tensor field on $E$ that is related to the curvature of $E$ by Einstein field equations. 

\end{itemize}

From ${\bf O11}$, the continuum approximation is valid in the large number limit of basic events and allows to match the pre-geometric structure with the ontological one. To prove ${\bf O11}$ as a theorem from more basic axioms is a major problem of the causal set approach to quantum gravity. I hope to discuss this issue elsewhere.

\begin{acknowledgements}
I thank Mario Bunge, Patrick D\"urr, Laurant Freidel, Santiago E. Perez-Bergliaffa, H. Vucetich, Janou Glaeser, Gerardo Primero, and Ferm{\'\i}n Huerta Mart{\'\i}n for stimulating discussions. Some parts of this work were presented in the XV Brazilian School on Gravitation and Cosmology and the international meeting GR 100. I thank Mario Novello for his kind invitations to deliver my lectures in such a stimulating environments. 

%If you'd like to thank anyone, place your comments here
%and remove the percent signs.
\end{acknowledgements}

% BibTeX users please use one of
%\bibliographystyle{spbasic}      % basic style, author-year citations
%\bibliographystyle{spmpsci}      % mathematics and physical sciences
%\bibliographystyle{spphys}       % APS-like style for physics
%\bibliography{}   % name your BibTeX data base

\begin{thebibliography}{99}

\bibitem{Birrell1982}
Broad, C.D. (1923). {\em Scientific Thought}. New York: Harcourt, Brace and Co.


\bibitem{Broad1923}
Birrell, N.D. and Davis, P.C.W. (1982). {\em Quantum Fields in Curved Space}. Cambridge: Cambridge University Press.


\bibitem{Bunge1967}
Bunge, M. (1967). {\itshape Foundations of Physics}. New York: Springer-Verlag.

\bibitem{Bunge1973}
Bunge, M (1973). \textit{Philosphy of Physics}. Dordrecht: Reidel.

\bibitem{Bunge1974a}
Bunge, M (1974a). \textit{Semantics I: Sense and Reference}. Dordrecht: Reidel.

\bibitem{Bunge1974b}
Bunge, M. (1974b). \textit{Semantics II: Meaning and Interpretation}. Dordrecht: Reidel. 


\bibitem{Bunge-77-Dordrecht}
Bunge, M. (1977). {\em Treatise of Basic Philosophy. Ontology I: The Furniture of the World}.
Dordrecht: Reidel.

\bibitem{Bunge1979}
M. Bunge (1979). {\itshape Causality in Modern Science}. New York: Dover, 2nd ed.

\bibitem{Bunge2003a}
Bunge , M. (2003a). {\itshape Philosophical Dictionary}. Prometheus Books, 2nd ed.

\bibitem{Bunge2003b}
Bunge , M. (2003b). {\itshape Emergence and Convergence}. Toronto: University of Toronto Press.

\bibitem{Bunge2006}
Bunge, M. (2006). {\itshape Chasing Reality: Strife over Realism}. Toronto: University of Toronto Press.


\bibitem{Covarrubias}
Covarrubias, G.M. (1993).
\newblock {An axiomatization of General Relativity}.
\newblock {\em International Journal of Theoretical Physics},  32, 2135-2154.

\bibitem{Craig-paper}
Craig, W.L (2008). The Metaphysics of Special Relativity: Three Views. In:  {\em Einstein, Relativity, and Absolute Simultaneity}, eds. Craig, W.L. and Smith, Q., 11-49. London: Routledge.


\bibitem{Craig}
Craig, W.L. and Smith, Q. , eds. (2008). {\em Einstein, Relativity, and Absolute Simultaneity}. London: Routledge.


\bibitem{Crisp-03-Oxford}
Crisp, T. (2003).
Presentism. In {\em The Oxford Handbook of Metaphysics}, eds. M. J. Loux \& D. W. Zimmerman, 211-245. Oxford: Oxford University Press.

\bibitem{Crisp-07-in}
Crisp, T. 2007.
Presentism, Eternalism and Relativity Physics. In {\em Einstein, Relativity and Absolute Simultaneity}, eds. W. L. Craig \& Q. Smith, 262-278. London: Routledge. 

\bibitem{Dowker}
Dowker, F. (2006).
\newblock {Causal sets as discrete specetime}.
\newblock {\em Contemporary Physics},  47, 1-9.


\bibitem{Eagleman}
Eagleman, D.M.  (2009). Brain time. In:  {\em What's Next: Dispatches from the Future of Science}, ed. M. Brockman, 155-169,  New York: Vintage Books.


\bibitem{Earman-Norton-1987}
Earman, J. and Norton, J.N.  (1987). What price substantivalism? The hole story.  {\em Brit. J. Phil. Sci.},  38, 515-525. 


\bibitem{Einstein-1907}
Einstein, A. (1907).  \"Uber das Relativit\"atsprinzip und die aus demselben gezogenen Folgerungen,
{\em Jahrbuch der Radioaktivit\"at}, 4,  411-462.

\bibitem{Einstein-1915}
Einstein, A. (1915). Die Feldgleichungen der Gravitation. {\em Preussische Akademie der Wissenschaften Berlin}, 844-847. 

\bibitem{Einstein-1916}
Einstein, A. (1916). N\"aherungsweise Integration der Feldgleichungen der Gravitation.  {\em Preussischen Akademie der Wissenschaften Berlin}. Part 1: 688–696.

\bibitem{Einstein-1918a}
Einstein, A. (1918a). Prinzipielles zur allgemeinen Relativitätstheorie. {\em Annalen der Physik}, 55, 241–244.


\bibitem{Einstein-1918b}
Einstein, A. (1918b). \"Uber Gravitationswellen. {\em Preussischen Akademie der Wissenschaften Berlin}. Part 1: 154–167.

\bibitem{Einstein-1920}
Einstein, A. (1920). Ether and the theory of relativity. In: Renn, J. and Schemmel, M.  (eds.),  (2007) {\em The Genesis of General Relativity, Vol. 3: Gravitation in the Twilight of Classical Physics}. Berlin: Springer, 613-619.

\bibitem{Ellis2010}
Ellis, G.F.R. and Rothman, T. (2010).
\newblock {Time and spacetime: The crystallizing Block Universe}.
\newblock {\em International Journal of Theoretical Physics},  49, 988-1003.


\bibitem{Graham2006}
Graham, D.W. (2006). {\em Explaining the Cosmos}. Princeton: Princeton University Press.

\bibitem{Folsing1998}
F\"olsing, A. (1998). {\em Albert Einstein}. New York: Penguin Books.

\bibitem{Friedman1983}
Friedman, M. (1983). {\em Foundations of Space-Time Theories}. Princeton: Princeton University Press.

\bibitem{Heller}
M. Heller (1990). {\itshape The Ontology of Physical Objects}. Cambridge: Cambridge University Press.

\bibitem{Hobson}
Hobson, H.P.,  Efstathiou, G.,  and Lasenby, A.N. (2006),  {\rm General Relativity}.  Cambridge: Cambridge University Press.

\bibitem{Hoefer-1996-JP}
Hoefer, C. (1996). The metaphysics of space-time substantivalism.  {\em The Journal of Philosophy}, 93, 5-27. 

\bibitem{Hoefer-1998}
Hoefer, C. (1998). Absolute versus relational spacetime: For better or worse, the debate goes on.  {\em Brit. J. Phil. Sci.},  49, 451-467. 

\bibitem{Jammer}
Jammer, M.  (2012). {\itshape Concepts of Space}. New York: Dover, 3rd ed.

\bibitem{Leibniz}
Leibniz, G.W. (2005).  {\it Discourse on Metaphysics and The Monadology}. Mineola, NY: Dover Publications.  

\bibitem{Leibniz2}
Leibniz, G.W. and  Clarke, S. (2000).  {\it Correspondence}, edited by R. Ariew. Indianapolis: Hackett Publishing Co..  

\bibitem{LePoidevin}
Le Poidevin, R. (2007).  {\it The Images of Time}. Oxford: Oxford University Press.

\bibitem{Malament1977}
Malament, D. (1977). The class of continuous timelike curves determines the topology
of spacetime. {\em Journal of Mathematical Physics}, 18, 1399-1404.

\bibitem{Mauldin1989}
Maudlin, T. (1989). The essence of spacetime. In: A. Fine and J. Leplin (eds.), {\em PSA} 1988, Volume 2, 82–91.


\bibitem{Mink}
H. Minkowski (1908).
\newblock Lecture ``Raum und Zeit, 80th Versammlung Deutscher Naturforscher (K$\ddot{o}$ln, 1908)'', {\em Physikalische Zeitschrift}, 10, 75-88 (1909).

\bibitem{Mellor} 
Mellor, M. J. (1998).  {\em Real Time II}. London: Routledge.


\bibitem{Zi-2011-PP} 
Mozersky, M. J. (2011). Presentism. In {\em The Oxford Handbook of Philosophy of Time}, ed. C. Callender, 122-144. Oxford: Oxford University Press.

\bibitem{Norton2014}
Norton, J.D. (2014). The Hole Argument, The Stanford Encyclopedia of Philosophy (Spring 2014 Edition), Edward N. Zalta (ed.), URL = $<$http://plato.stanford.edu/archives/spr2014/entries/spacetime-holearg/$>$.

\bibitem{Oaklander} 
Oaklander, L.N. (2004).  {\em The Ontology of Time}. Amherst, NY: Prometheus Books.

\bibitem{Oriti}
Oriti, D. (2014).
\newblock {Disappearance and emergence of space and time in quantum gravity}.
\newblock {\em Studies in the Histroy and Philosophy of Modern Physics}, 46, 186-199.


\bibitem{Bergliaffa1}
Perez-Bergliaffa, S.E., Romero, G.E., and Vucetich, H. (1993).
\newblock {Axiomatic foundations of nonrelativistic Quantum Mechanics: A realistic approach}.
\newblock {\em International Journal of Theoretical Physics},  32, 1507-1522.

\bibitem{Bergliaffa3}
Perez-Bergliaffa, S.E., Romero, G.E., and Vucetich, H. (1998).
\newblock {Toward an axiomatic pregeometry of space-time}.
\newblock {\em International Journal of Theoretical Physics},  37, 2281-2298.

\bibitem{Peterson-Silberstein}
Peterson, D, and Silberstein, M. 2010. In: {\em Space, Time, and Spacetime: Physical and Philosophical Implications of Minkowski's Unification of Space and Time}, ed. V. Petkov, 209-237. Berlin, Heidelberg: Springer.

\bibitem{Petkov2006} 
Petkov, V. (2006). Is there an alternative to the Block Universe view?.  In:  {\em The Ontology of Spacetime}, ed. D. Dieks, 2007-228. Utrecht: Elsevier. 

\bibitem{Poeppel}
P\"oppel, E. (1988).  {\it Mindworks. Time and Conscious Experience}. Orlando, Fl: HBJ Publishers. 

\bibitem{Poepper}
Popper,  K. (2005).  {\it Unended Quest}. NY: Taylor and Francis e-Library.  

\bibitem{Put-1967-JP}
Putnam, H. (1967). Time and physical Geometry. {\em The Journal of Philosophy},  64, 240-247. 

\bibitem{Romero2012}
Romero, G.E. (2012).
\newblock {Parmenides reloaded}.
\newblock {\em Foundations of Science},  17, 291-299.

\bibitem{Romero2013a}
Romero, G.E. (2013a). From change to spacetime: an Eleatic journey,
{\em Foundations of Science},  18, 139-148.


\bibitem{Romero2013b}
Romero, G.E. (2013b). Adversus singularitates: The ontology of space-time singularities, {\em Foundations of Science} , 18 297-306.


\bibitem{Romero2014-ont-GR} Romero, G.E. (2014). The ontology of General Relativity,  in M. Novello, S.E. Perez Bergliaffa, {\em Gravitation and Cosmology}, Chapter 8, Cambridge, Cambridge Scientific Publishers, Ltd, pp. 1-15 .

\bibitem{Romero2015}
Romero, G.E. (2015). Present time,
{\em Foundations of Science},  20, 135-145.


\bibitem{Romero2014book} Romero, G.E., \& Vila, G.S. (2014). {\it Introduction to Black Hole Astrophysics}. Heidelberg: Springer.

\bibitem{Romero2014-BH-pres} Romero, G.E., \& P\'erez, D. (2014). Presentism meets black holes.  {\em European Journal for Philosophy of Science}, 4, 293-308

\bibitem{Rovelli} Rovelli, C. (2004). {\it Quantum Gravity}. Cambridge: Cambridge University Press. 

\bibitem{Rynasiewicz}  Rynasiewicz, R. (1996). Absolute versus Relational Spacetime: An Outmoded
Debate?. {\emph The Journal of Philosophy},  93,  279-306.

\bibitem{Sau-02-Ca}
Saunders, S. (2002). How relativity contradicts presentism. In: {\em Time, Reality $\&$ Experience}, Royal Institute of Philosophy, Supplement, ed. C. Callender, 277-292. Cambridge, New York: Cambridge University Press. 


\bibitem{Smart-63-Ca}
Smart, J. J. C. (1963). {\em Philosophy and Scientific Realism}. London: Routledge.

\bibitem{Smeenk}
Smeenk, C. (2014). Einstein's role in the creation of relativistic cosmology. In: {\em The Cambridge Companion to Einstein}, eds. Janssen, M. and Lehner, C., 228-269. Cambridge: Cambridge University Press.

\bibitem{Sorabji1983}
Sorabji, R. (1983). {\em Time, Creation and the Continuum}. Chicago: The Chicago University Press. 

\bibitem{Stein-1968-JP}
Stein, H. 1968. On Einstein-Minkowski Space-Time. {\em Journal of Philosophy}, 65, 5-23. 

\bibitem{Taylor} 
Taylor, J. H., and  Weisberg, J. M. (1982). A new test of general relativity - Gravitational radiation and the binary pulsar PSR 1913+16 . {\em The Astrophysical Journal}, 253, 908–920.

\bibitem{Wuthrich2010}
W\"uthrich, C. (2010). No Presentism in Quantum Gravity. In: {\em Space, Time, and Spacetime: Physical and Philosophical Implications of Minkowski's Unification of Space and Time}, ed. V. Petkov, 257-258. Berlin, Heidelberg: Springer.

\end{thebibliography}

% Non-BibTeX users please use
%\bibliographystyle{aipproc}   % if natbib is available
%\bibliographystyle{aipprocl} % if natbib is missing

%%%%%%%%%%%%%%%%%%%%%%%%%%%%%%%%%%%%%%%%%%%
%% You probably want to use your own bibtex database here
%%%%%%%%%%%%%%%%%%%%%%%%%%%%%%%%%%%%%%%%%%%

\newpage

\section*{Gustavo E. Romero} Full Professor of Relativistic Astrophysics at the University of La Plata and Superior Researcher of the National Research Council of Argentina. A former President of the Argentine Astronomical Society, he has published more than 350 papers on astrophysics, gravitation and the foundations of physics. Dr. Romero has authored or edited 10 books (including {\sl Introduction to Black Hole Astrophysics}, with G.S. Vila, Springer, 2014). His main current interests are on high-energy astrophysics, black hole physics, and ontological problems of spacetime theories.

\end{document}